\documentstyle[12pt]{article}

\def\be{\begin{equation}}
\def\ee{\end{equation}}
\def\bea{\begin{eqnarray}}
\def\eea{\end{eqnarray}}
\def\<{\langle}
\def\>{\rangle}
\newcommand{\CR}{\nonumber \\}
\newcommand{\pa}{\partial}
\newcommand{\cL}{{\cal L}}
\newcommand{\th}{\theta}
\newcommand{\A}{\alpha}
\newcommand{\ph}{\phi}
\newcommand{\phd}{\phi^\dagger}
\newcommand{\ww}{w}

\newcommand{\rf}[1]{(\ref{#1})}

\renewcommand{\l}{\lambda}
\renewcommand{\L}{\Lambda}
\renewcommand{\b}{\beta}
\renewcommand{\a}{\alpha}
\newcommand{\n}{\nu}
\newcommand{\m}{\mu}

\newcommand{\sg}{\sigma}

\newcommand{\oq}{\frac{1}{4}}
\newcommand{\dg}{\dagger}
\newcommand{\tr}{{\rm Tr}\,}
\newcommand{\ra}{\right\rangle}
\newcommand{\la}{\left\langle}

\def\void{}
\def\labelmark{}

\newenvironment{formula}[1]{\def\labelname{#1}
\ifx\void\labelname\def\junk{\begin{displaymath}}
\else\def\junk{\begin{equation}\label{\labelname}}\fi\junk}%
{\ifx\void\labelname\def\junk{\end{displaymath}}
\else\def\junk{\end{equation}}\fi\junk\labelmark\def\labelname{}}

{\ifx\void\labelname\def\junk{\end{array}\end{displaymath}}
\else\def\junk{\end{array}\right.\end{equation}}
\fi\junk\labelmark\def\labelname{}\def\junk{}
}

\newcommand{\beq}{\begin{formula}}
\newcommand{\eeq}{\end{formula}}
\newcommand{\beqv}{\begin{formula}{}}

\begin{document}

\begin{flushright} 
 NBI-HE-97-22
 
 UB-ECM-PF 97/13

 July 1997
\end{flushright}
 
\renewcommand{\thefootnote}{\fnsymbol{footnote}}

\begin{center}
\vspace{24pt}
{ \large \bf $U(1)$ lattice gauge theory and \\
$N=2$ supersymmetric Yang-Mills theory }

\vspace{24pt}

{\sl Jan Ambj\o rn$^{a,}$\footnote{E-mail: ambjorn@nbi.dk}, 
Dom\`enec Espriu$^{b,}$\footnote{E-mail: espriu@greta.ecm.ub.es}} and 
{\sl Naoki Sasakura$^{a,}$\footnote{E-mail: sasakura@nbi.dk}}

\vspace{24pt}
$^a$The Niels Bohr Institute,  \\
University of Copenhagen,\\ 
Blegdamsvej 17, \\
DK-2100 Copenhagen \O ,\\
 Denmark \\

\vspace{18pt}

$^b$Department of Physics, \\
University of Barcelona, \\
Diagonal 647, \\
E-08028, Barcelona, \\
Spain

\end{center}

\renewcommand{\thefootnote}{\fnsymbol{footnote}}
\setcounter{footnote}{0}

\vfill

\begin{center}
{\bf Abstract}
\end{center}

\vspace{12pt}

\noindent
We discuss the physics of four-dimensional compact 
$U(1)$ lattice gauge theory from the point 
of view of softly broken $N$=2 supersymmetric $SU(2)$ Yang-Mills theory.
We provide arguments in favor of (pseudo-)critical mass exponents
1/3, 5/11 and 1/2, in agreement with the values observed in the 
computer simulations. We also show that the $J^{CP}$ assignment of 
some of the lowest lying states can be naturally explained. 

\vfill

\newpage

\section{Introduction}\label{sec1}

Recent computer simulations indicate that compact $U(1)$ lattice gauge
theory 
has a second order phase transition for a finite value of $\b$
and that the critical exponents associated with the transition are 
non-trivial \cite{jln,NEU,rebbi,More}. 
This is a remarkable situation since it, according to 
ordinary folklore, implies that there exists a non-trivial continuum 
field theory with these critical exponents. The lattice system 
itself is fairly simple and one can transform it to a dual gauge field
coupled to a monopole field \cite{polyakov,bmk}. 
The monopole field is a lattice
artifact caused by the compactness of the $U(1)$ gauge group 
used on the lattice, in the same way  as the topological defects 
of the $XY$-model in two dimensions are caused by the $O(2)$ symmetry.
The $XY$ phase transition is related to the ``liberation'' of these
topological defects and is of infinite order. Associated with
this transition we have an Euclidean quantum field theory, namely 
the sine-Gordon theory for a special value of the coupling constant.
The $U(1)$ lattice gauge theory in three dimensions describes formally
a gauge theory with lattice monopoles. This theory has no other
fixed point than the gaussian one, where we recover the familiar 
three-dimensional electrodynamics and which has no monopoles at all. 
There is no 
way of taking the continuum limit in which the monopoles survive. 
However, there exists 
a three-dimensional non-Abelian $SU(2)$ gauge-Higgs theory, the 
Georgi-Glashow model, which has monopoles with finite action\footnote{In 
fact, the monopoles act as instantons in the three-dimensional Euclidean 
theory.} and where 
the $SU(2)$ gauge theory is spontaneously broken to $U(1)$. The physics 
of this model, confinement of the $U(1)$ charge and  a corresponding 
non-vanishing string tension and a massive dual photon, is qualitatively 
the same as in the three-dimensional compact $U(1)$ lattice gauge 
theory \cite{polyakov},
and the lattice theory 
describes similar long distance physics as the continuum model,
but it cannot be used to {\it define} 
in a rigorous way the full quantum field theory by approaching a 
fixed point.

In the case of compact $U(1)$ in four dimensions, the situation is,
at least apparently, more like the the $XY$-model 
in two dimensions than like the the compact $U(1)$ theory in three 
dimensions: we have a critical point, a phase transition which may 
be of an order higher than one, and (measured) non-trivial exponents.
However, 
in the case of the $XY$-model it was a non-trivial task to identify an 
underlying local
quantum field theory, and in the case of the $U(1)$ lattice 
theory in four dimensions we have, frankly speaking, no obvious
candidate 
at all. We can thus take a pragmatic 
attitude
and simply ask whether there exists a continuum model field theory 
which has qualitatively the same features as observed for the 
$U(1)$ lattice theory. Such a question has a meaning 
even if it may turn out that the $U(1)$-transition 
is a (weakly) {\em first order phase transition}\footnote{The history of
the 
$U(1)$ phase transition is rather turbulent. First it was classified as 
a second order transition, with critical exponents fluctuating from 
mean field exponents to non-trivial exponents. Then hysteresis was 
discovered and extensive computer simulations pointed to a first order 
transition.  Finally new computer simulations, either 
changing topology of space-time from toroidal to spherical \cite{jln} 
or adding by hand monopole-like terms \cite{rebbi}, 
restored the second order transition, 
again with non-trivial exponents. See \cite{NEU} for more references.}, 
since in this case the $U(1)$ 
lattice theory and the underlying continuum theory should still possess 
analogous long distance properties:
a phase where the $U(1)$ charge is confined, 
where the corresponding string tension is non-zero and where a number 
of ``gauge balls'' are observed,
and a Coulomb phase where the monopoles anti-screen. Approaching the 
phase transition the non-trivial (pseudo-)scaling of the lattice model
should 
be present in the continuum model too, and finally, some universality 
features of the renormalized charge in the  $U(1)$-lattice 
theory \cite{CAR,luck}
should also be explained. It is clear that this is a non-trivial 
task for the continuum quantum theory, irrespectively of whether or not
a genuine scaling limit can be defined for the lattice model.

An obvious candidate for a  
continuum field theory which may describe the same physics as the $U(1)$ 
lattice gauge theory is a softly broken $N=2$ supersymmetric $SU(2)$ 
gauge theory. Before the soft breaking it describes at low energies
a $U(1)$ theory which for certain values of the moduli of the theory 
consists of  a light monopole hyper-multiplet interacting with a dual 
photon multiplet \cite{SEIWIT}. 
After  breaking to $N\!\!=\!1$ supersymmetry it describes, at low
energies, in 
the vicinity of the same value of the moduli, a 
$U(1)$ theory with a massive dual photon, non-zero string tension and a 
monopole condensate \cite{SEIWIT}. 
In order to make contact to the lattice $U(1)$ theory 
we have to induce further soft breaking of the $N\!\!=\!1$
supersymmetry.
This is not under control to the same extent as the breaking from 
$N=2$ to $N\!\!=\!1$. In the rest of this article we try to argue,
mainly  heuristically,  that there exists a softly broken 
$N=2$ supersymmetric theory which describes most of the features 
observed for the $U(1)$ lattice gauge theory, such that this 
lattice theory may play the same role relative to certain softly 
broken $N=2$ supersymmetric gauge theories as compact three-dimensional 
$U(1)$ lattice gauge theory plays with respect to the Georgi-Glashow
model.

The rest of the article is organized as follows: in section \ref{sec2}
we review shortly the results from the numerical simulations of the 
four-dimensional compact $U(1)$ lattice gauge theory. 
In section \ref{sec3} we describe the soft breaking of 
the $N=2$ supersymmetric gauge theory relevant for the 
phase transition between a confining $U(1)$ theory and the $U(1)$ theory 
in the Coulomb phase. 
In section \ref{sec4} we try to match the physics of section \ref{sec2}
and \ref{sec3}. Finally, section \ref{sec5} contains a discussion.

\section{Compact U(1) lattice gauge theory}\label{sec2}

The lattice theory is defined by the action 
\beq{*0}
S= - \sum_{\Box} \b \cos \th_{\Box},
\eeq
where the summation is over all plaquettes $\Box$ on the lattice and 
$\th_\Box \in [0,2\pi)$ is the argument of the product of the $U(1)$
link 
variables around the plaquette $\Box$. If we define $F_{\m\n}$ by 
$\th_\Box = a^2 e_0 F_{\m\n}$, where $a$ is the lattice spacing and 
$\b = 1/e_0^2$, the action \rf{*0} reduces in 
formal limit $a \to 0$ to the standard continuum expression 
$S = \oq \int d^4x F_{\m\n}^2$. It is possible to add additional 
terms to \rf{*0} to get an extended coupling constant space. 
Many of the computer simulations are performed in this extended 
coupling constant space. We refer to \cite{NEU} for details.

Part of the physics of the $U(1)$ lattice gauge theory is well
understood
\cite{polyakov,bmk}. It has  a two phase structure. For large bare 
coupling constant $e_0$ the system exhibits confinement of electric
charge,
while  it for small values of the bare coupling is in the Coulomb phase 
with a massless photon. As the coupling constant increases, the coupling 
to diluted topological excitations, which can be interpreted as
(lattice)
magnetic monopole loops, decreases, and these magnetic monopole loops 
unbind at the phase transition point, beyond which the (lattice)
monopoles
condense and cause confinement of the electric charge and a linear
rising
potential between static test charges by the 
dual Meissner effect. In the Coulomb phase the static charges interact 
via the Coulomb potential, and when the bare charge approaches the
critical 
value $e_0^c$ the monopole loops will renormalize (anti-screen) the
charge
according to 
\beq{*1}
V_{coulomb} (r) = - \frac{\a_r}{r}, ~~~~\a_r \equiv \frac{e^2_r}{4\pi},
\eeq
where the relation between the renormalized and the bare charge 
has been conjectured to be \cite{CAR,luck} 
\beq{*2}
\a_r(\a_0) = \a_r^c - {\rm const.}\, \left( 1
-\frac{\a_0}{\a_0^c}\right)^\l,
\eeq
with both $\a_r^c$ and $\l$ universal. While $\l$ appears as a 
standard critical exponent, it is more surprising that $\a_r^c$ should 
be universal. The last conjecture arose by analogy with the $XY$-model
and 
\rf{*2} is consistent with present numerical evidence. 

The recent numerical simulations have gone much further \cite{NEU}. The 
masses of so-called ``gauge balls''  have been measured, 
and approaching the critical point 
{}from the confinement region it is found that the masses $m_j(\b)$
and the square root of the string tension, $\sqrt{\sg(\b)}$ scale as 
\beq{*3}
m_j (\b) \sim c_j (\b_c -\b)^{\n},~~~\sqrt{\sg(\b)} \sim (\b_c
-\b)^{\n},
~~~~~\n \simeq 0.35 \pm 0.03
\eeq
For most of the masses $m_j$ there exists in addition definite spin,
parity and charge conjugation assignment $J^{PC}$. We refer to
\cite{NEU}
for details.

The only exception from the above scaling 
is a state $J^{PC}= 0^{++}$ which is observed to 
scale with Gaussian exponent:
\beq{*4}
m_{0^{++}} \sim (\b_c -\b)^{\n_g},~~~~\n_g \simeq 0.51 \pm 0.03
\eeq

\vspace{12pt}
Another series of recent simulations use the lattice action \rf{*0},
but add by hand a lattice monopole term \cite{rebbi}, such that  
\beq{*5}
S_{mono} = - \sum_{\Box} \b \cos \th_{\Box} -\l \sum_{\rho,x}
|M_{\rho,x}|, 
\eeq
where $M_{\rho,x}=\varepsilon_{\rho\sigma\mu\nu}
(\bar \theta_{\mu\nu,x+\sigma}-\bar \theta_{\mu\nu,x})/4\pi$ and the
physical
flux $\bar \theta_{\mu\nu,x} \in [-\pi,\pi)$ is related to the plaquette 
angle $\theta_{\mu\nu,x}\in (-4\pi,4\pi)$ by $\theta_{\mu\nu,x}=
\bar \theta_{\mu\nu,x}+2\pi n_{\mu\nu,x}$.
According to \cite{rebbi}, a sufficiently large fixed  value of the 
coupling constant $\l$ ensures a second order transition for a 
finite value $\b_c(\l)$, and with an exponent $\n_{mono}$,
which is not the mean field exponent \rf{*4} nor the non-trivial 
exponent in eq.\ \rf{*3}. According to \cite{rebbi} one has 
\beq{*6}
\n_{mono} \simeq 0.44 \pm 0.02.
\eeq

The question we ask is simply: is there any continuum quantum 
field theory which is compatible with (some of) the above mentioned
lattice measurements ?

\section{U(1) from Supersymmetric  Yang-Mills theory}\label{sec3}

In this section we discuss  briefly the Seiberg-Witten derivation 
of a low energy effective $U(1)$ theory containing monopoles
{}from the $N\!\!=\!2$ supersymmetric $SU(2)$ 
theory \cite{SEIWIT}, 
and how soft breaking and the addition of a certain superpotential 
can create a $U(1)$ confinement-deconfinement phase transition lying 
entirely in the $N=0$ sector,   
with physics resembling those outlined above for lattice $U(1)$.

\subsection{The Seiberg-Witten Solution and its soft breaking}

While the quantum fluctuations of $N\!\!=\!4$ 
supersymmetric Yang-Mills theory are 
trivial, those of $N\!\!=\!2$ supersymmetric Yang-Mills 
theory consist of an infinite
series of instanton corrections as well as a one-loop contribution
\cite{SEI}. 
Thus $N\!\!=\!2$ supersymmetric Yang-Mills theory contains non-trivial
physics,
but the symmetry constrains the quantum fluctuations so much that 
the theory can be analyzed in considerable detail. 
The ground state is 
parameterized by an order parameter $u= \la \tr \phi^2 \ra$
corresponding 
to the breaking of $SU(2)$ to $U(1)$.  For large values of $u$ we have 
a standard scenario: at energy scales  $\mu\gg\sqrt{u}$ all 
field theoretical degrees of freedom contribute to the $\b$-function,
which corresponds to the asymptotically free theory. For  energies
lower than $\sqrt{u}$ only the $U(1)$ part of the theory is effective.
In these considerations the dynamical confinement scale 
$$\L_{N=2}^4= \mu^4\exp(-8\pi^2/g(\mu)^2)$$ 
obtained by the one-loop perturbative calculation plays no role. 
The remarkable observation by Seiberg and Witten was that even 
when $u \leq \L^2_{N=2}$, where one would naively 
expect that non-Abelian dynamics was 
important, the system remains in  the $U(1)$ Coulomb phase due 
to supersymmetric cancellations of non-Abelian quantum fluctuations. 
As $u$ decreases the effective electric charge  associated
with the unbroken $U(1)$ part of $SU(2)$ increases, while the masses of
the solitonic excitations which are present in the theory, will
decrease.
Dictated by monodromy properties of the so-called prepotential of the 
effective low energy Lagrangian, the monopoles become massless at 
a point $u \sim \L^2_{N=2}$ where the effective electric charge
has an infrared Landau pole and diverges.
However, in the neighborhood of $u \sim \L^2_{N=2}$, 
this strongly coupled theory has an effective Lagrangian 
description as a weakly coupled theory when expressed  
in terms of dual variables, namely a monopole hyper-multiplet  
and a dual photon vector multiplet. The perturbative coupling constant
is now $g_D=4\pi/g$ and the point where monopoles condensate corresponds
to
$g_D=0$.

Another remarkable observation of Seiberg and Witten is that 
the breaking of $N\!\!=\!2$ to $N\!\!=\!1$ supersymmetry 
by adding a  mass term  superpotential will generate 
a mass gap, originating from a condensation of the monopoles.
By the dual Meissner effect this theory confines the electric 
$U(1)$ charge at distances larger than the inverse $N\!\!=\!2$
symmetry breaking scale. In terms of the underlying microscopic theory
it is 
believed that the reduction of symmetry from $N\!\!=\!2$ to  $N\!\!=\!1$ 
allows excitations closer to generic non-supersymmetric 
``confinement excitations'', but that the soft breaking ensures
that the theory is still close enough to the $N\!\!=\!2$ to remain an
effective $U(1)$ theory\footnote{Or, as many people believe, 
that the generic confinement excitations {\it are} ``monopole-like'', 
as described in the softly broken $N\!\!=\!2$ theory, and that 
the difference in the allowed fluctuations in ordinary $QCD$ 
compared to softly broken $N\!\!=\!2$ are such that they do not 
modify the confinement mechanism. Whether this is true or not true 
will be irrelevant for our use of the Seiberg-Witten results.}.

Thus we see that in the Seiberg-Witten scenario we can, by introducing 
the mass term superpotential,
describe a $U(1)$ confining-deconfining transition. 
However, the transition is from the $N\!\!=\!2$ Coulomb phase to the 
$N\!\!=\!1$ confining phase, while we are interested in a $U(1)$
confining-deconfining transition occuring in the  $N\!\!=\!0$ sector.

The breaking down to $N\!=\!0$ has been analyzed in a number of 
papers \cite{EVA,ALV}, where soft breaking via 
spurion fields of  $N\!\!=\!1$ 
and $N\!\!=\!2$ supersymmetric gauge theories are discussed (see also 
\cite{SEE}).
One of the motivations behind these studies has been to test if the 
Seiberg-Witten scenario can be extrapolated to realistic 
models for $QCD$ confinement. Here we will use 
the same philosophy to test if it is possible to 
explain the lattice physics described in section \ref{sec2}
in terms of continuum physics of an extended model, precisely 
as the physics of compact lattice $U(1)$ in three dimensions 
is described by the continuum Georgi-Glashow model.  

\pagebreak

\subsection{The model}

\subsubsection{Unbroken $N=2$}

The starting point is the $N\!\!=\!2$ supersymmetric Yang-Mills theory.
In $N\!\!=\!1$ superspace notation the bare Lagrangian is given by
\beq{barlag}
{\cal L}_{bare}=\frac2{g^2} \int d^2 \th d^2 \bar{\th} \;
{\rm Tr}\,(\Phi^\dagger e^{-2V}\Phi e^{2V})
+\frac1{2g^2}\left(\int d^2\th \;{\rm Tr\,}W^\A W_\A + h.c. \right),
\eeq
where all fields are in the fundamental representation, where $g$ is the 
bare gauge coupling, and where the $N\!\!=\!1$ chiral 
multiplet\footnote{Our superfield notation is as follows:
$\Phi = a_\Phi+ \sqrt{2}\th \psi_\Phi + \th \th f_\Phi$ with 
the $F$-term being the coefficient to $\th\th$, $\Phi|_F \equiv f_\Phi$, 
the $A$-term the coefficient to 1, $\Phi|_A \equiv a_\Phi$, and
the $D$-term the coefficient to $\th^2 \bar{\th}^2$, i.e. for 
instance  $\Phi^\dg\Phi|_D \equiv |f_\Phi|^2$, etc.} $\Phi=(\phi,\psi)$
and the $N\!\!=\!1$ vector multiplet $W_\A=(v_\mu,\lambda)$ 
constitute the $N\!\!=\!2$ vector multiplet in the Wess-Zumino gauge.

The elimination of the $D$--component in \rf{barlag} 
produces the term $ {\rm Tr\,}([\ph,\phd]^2)/g^2$ in the potential,
which has a flat direction $\ph=a\sigma^3/2$ with an arbitrary complex
number 
$a$. The order parameter is given by $u=\langle {\rm
Tr\,}(\ph^2)\rangle$. 
For generic values of $u$, the $SU(2)$ gauge 
symmetry breaks down to $U(1)$ and 
the low energy effective theory will be given by the Lagrangian of the 
$N\!\!=\!2$ $U(1)$ supersymmetric Yang-Mills theory  
\beq{seilag}
{\cal L}_{SW}=
\frac{1}{4\pi}{\rm Im}\left[ \int d^2\th d^2\bar{\th}\; 
\frac{\pa {\cal F}}{\pa A}\,\bar{A}
+\frac12 \int d^2 \th \;\frac{\pa^2{\cal F}}{\pa A^2}\,W^\A W_\A\right],
\eeq
where the prepotential ${\cal F}$ is a holomorphic function of $A$
(whose
lowest component is $a$) and 
the dynamical scale $\Lambda_{N=2}$, generated by the 
non-Abelian interactions. 

The functional form of the prepotential is uniquely determined from the 
monodromy properties
of the singularities expected from duality and the spectrum of
solitonic states \cite{SEIWIT,ORA}. 
In the $SU(2)$ case, there are three singularities. One is
the semi-classical one with $a$ at infinity, and the others at, say
$u=\pm \Lambda^2_{N=2}$, 
where a monopole or a dyon becomes massless, respectively.
Near the point $u=\L^2_{N=2}$  where the monopoles become massless and
the 
theory is strongly coupled in terms of the original field variables, 
a duality transformation results in a weakly coupled effective
Lagrangian of the monopole field 
\be
{\cal L}_M=\int d^2\th d^2\bar{\th} \;
(M^\star e^{2V_D}M+\tilde{M}^\star e^{-2V_D} 
\tilde{M})+\left( \int d^2\th \; \sqrt{2} A_DM\tilde{M}+ h.c.\right),
\label{monlag}
\ee
where $(M,\tilde{M})$ is the monopole hyper-multiplet and where  
$A_D$ and $V_D$ represent the $N\!\!=\!2$ vector multiplet of the dual
photon.

As mentioned, Seiberg and Witten introduced a breaking of the 
$N\!\!=\!2$ supersymmetry to $N\!\!=\!1$ by adding a 
superpotential  $b \,U$, $U =\tr \Phi^2$,
for the $N\!\!=\!1$ chiral multiplet. Adding this piece
breaks the flatness of the scalar potential and $u$ is no longer
a free parameter.
As long as the breaking parameter $b$ is 
suitably small, by comparison to $ \L_{N=2}$, the total 
superpotential is obtained as the sum of $b \,U$
and the potential from \rf{monlag}, i.e.\
\beq{suppot}
W = \sqrt{2}A_D M \tilde{M} + b\, U(A_D).
\eeq
When the breaking parameter $b \neq 0$, the vacuum, defined by $dW =0$,
satisfies
\bea
\sqrt{2} m \tilde{m}+ b \frac{d u}{da_D} &=&0, \CR
a_Dm=a_D\tilde{m}&=&0,
\label{suppot1}
\eea
where $m$, $\tilde m$ and $a_D$ denote the corresponding scalar
components
of $M$, $\tilde M$ and $A_D$, respectively. 
The solution is
\bea
m &=& \tilde{m}= \sqrt{-b u'(0)/\sqrt{2}} \sim \sqrt{b \L_{N=2}}, \CR
a_D &=& 0.
\label{suppot2}
\eea
The interpretation of \rf{suppot2} is that a monopole condensate is
formed
and that the potential generated from the superpotential \rf{suppot} in
this 
respect behaves like an ordinary Ginzburg-Landau potential.    

\subsubsection{Soft breaking by spurions}

A more general scheme of soft breaking 
of the Seiberg-Witten solution,  still respecting  
the monodromy properties of the singularities, was 
obtained in the references \cite{ALV}. 
They introduced a spurion $N\!\!=\!2$ vector multiplet, the dilaton
spurion, 
and the dynamical scale\footnote{In the rest of this section
we drop the suffix $N$=2 on $\L$ in order to avoid too cumbersome a
notation.}
$\Lambda$ is expressed as $\exp(is)$, where $s$
denotes the lowest scalar component of the dilaton spurion $S$.  
Thus the effective Lagrangian of the softly broken $N\!\!=\!2$
supersymmetric 
Yang-Mills theory is given by
\bea
{\cal L}_{soft}&=&
\frac{1}{4\pi}{\rm Im}\left[ \int d^2\th d^2\bar{\th}\; 
\frac{\pa {\cal F}}{\pa A^i}\bar{A}^i
+\frac12 \int d^2 \th \;\frac{\pa^2{\cal F}}{\pa A^i\pa A^j}\,{W^i}^\A 
{W^j}_\A \right],\CR
&&\ \ i=0,1;\ \ A^0=S,\ A^1=A,
\label{brolag}
\eea
where the lowest component and the auxiliary fields of the spurion
fields 
$A^0$ and $W_\A^0$ are frozen to constant values, thus breaking 
the $N\!\!=\!2$ directly down to $N\!\!=\!0$. 

In reference \cite{ALV}, the softly broken model with 
${\cal L}_{soft}+{\cal L}_M$ given by (\ref{monlag}) and (\ref{brolag})
is
analyzed. Since the monopoles have already been included in the heavy
modes
which are integrated out in ${\cal L}_{SW}$, the authors of \cite{ALV} 
argue that the monopole field in \rf{monlag} should be understood as 
representing the classical monopole field. 
This softly broken model was shown to 
be in the confinement phase, in the same way as the 
original $N\!\!=\!1$ model of Seiberg and Witten,  the dynamics of the 
confinement being monitored by a monopole condensation dictated 
by the freezing of $S$ (see \cite{ALV} for details). In the approach
of these authors, the parameter $F_0$, which
controls the breaking down to $N\!=\! 0$, plays the same role
as the parameter $b$ in the approach of Seiberg and Witten, and
has the same undesirable  (for our purposes of comparing with 
compact $U(1)$ on the lattice) feature
that there is no $N\!\!=\!0$ Coulomb phase.

\subsubsection{Monitoring confinement-deconfinement}

In order to create a scenario which is closer 
to the confinement-deconfinement transition observed on the lattice 
we need to have an $N\! = \! 0$ theory on both sides
of the confining transition. 

To achieve this purpose we add an additional $N=1$ Lagrangian
\be
{\cal L}_z=\int d^2\th d^2\bar{\th} \;z^\dagger z + \left(\int d^2 \th\
l\,z
\left(\ww-{\rm Tr\,}(\Phi^2)\right) + h.c.\right). 
\label{zetlag}
\ee
to the original Lagrangian (\ref{barlag}).
Here $z$ is an $N\!\!=\!1$ chiral multiplet without any gauge charges,
and $l$ and $w$ are free complex parameters.
In the $l\rightarrow 0$ limit, $z$ decouples from the original system
and will go back to the model mentioned  in the previous subsection.
In the $l\rightarrow \infty$ limit, the kinetic term of $z$ is
negligible, and
$z$ is an auxiliary field.

Consider the $N\!\!=\!1$ case of ${\cal L}_{bare}+{\cal L}_z$ 
in (\ref{barlag}) and (\ref{zetlag}).
After elimination of the $D$ and $F$ components, we obtain the potential
\be
V_{bare+z}=|l|^2|\ww-{\rm Tr\,}\ph^2|^2+4g^2|lz|^2{\rm Tr\,}(\phd\ph)
+\frac1{g^2}{\rm Tr\,}([\ph,\phd]^2).
\label{potbarz}
\ee 
The first term of (\ref{potbarz}) will
constrain the
value of  ${\rm Tr\,}\,\ph^2$ to be close to $\ww$; when $\ww$
is very large the system can be
treated 
semi-classically. Thus the gauge group will be broken at the scale 
of order $\sqrt{\ww}$ and the system will be in the $U(1)$ Coulomb
phase.
Taking the value $\ww$ smaller, the effective coupling of the system 
becomes stronger, but precisely as for the 
original model \rf{barlag}, the holomorphy argument \cite{HOL} ensures 
that the system will stay in the $U(1)$ Coulomb phase.  
On the other hand, 
if the soft-breaking terms are introduced and the supersymmetry is
broken, 
the delicate cancellation will be lost and the system will naturally be
in 
the confinement phase for small $\ww$.

Now consider the effect of the addition of ${\cal L}_z$ in the dual
picture.
This system is naively expected to be described by the effective
Lagrangian
${{\cal L}_{SW}}^D+{\cal L}_M+{\cal L}_z$, where ${{\cal L}_{SW}}^D$ is
the counterpart of ${\cal L}_{SW}$ in the dual variables.
But, since this is the effective low energy lagrangian after the
breakdown 
from $SU(2)$ to $U(1)$ and ${\cal L}_z$ respects only $N=1$
supersymmetry, 
some quantum corrections can appear.
To find the possible quantum corrections to the superpotential, consider
the global symmetry $U(1)\times U(1)_R$ of this system: 
$W^\A W_\A = (0,-2),\ A = (1,0)$. The charges of the fields in the dual
picture and the couplings are given by 
\bea
{W_D}^\A {W_D}_\A = (0,-2), & M\tilde{M} = (-1,-2), & A_D = (1,0), \CR
\Lambda = (1,0),    &zl = (-2,-2),          &  \ww = (2,0).
\label{cha}
\eea
The first $U(1)$ is anomalous, but the anomaly is cancelled by assigning
a charge to the dynamical scale $\Lambda$ \cite{HOL} 
as in the case where $\cL_z$ is absent. 
Based on the charge assignments \rf{cha}
the corrections to the superpotential conserving the charges must 
have the form 
\be
W_{q.c.}= ({W_D}^\A {W_D}_\A)^{1+n_1} (M\tilde{M})^{n_2} (zl)^{-n_1-n_2}
\Lambda^{-2n_1-n_2} f\left(\frac{A_D^2}\ww,\frac{\Lambda^2}{\ww} \right)
\ee
with integers $n_1\geq -1$ and $n_2\geq 0$ and a two-parameter function
$f$.
In this derivation we assumed that the generated superpotential is
regular
at  ${W_D}^\A,M,\tilde{M} \sim 0$. 
In the $l\rightarrow 0$ limit, $z$ decouples and $W_{q.c.}$ should
vanish.
This requirement determines that $n_1=-1$ and $n_2=0$.  
Moreover the parameter $\Lambda$ would  
appear like $\Lambda^4$ in the instanton corrections to the
superpotential. 
Thus we can parameterize the quantum corrections to the superpotential
as
\be
W_{q.c.} = \frac{\Lambda^4 zl}\ww f\left(\frac{A_D^2}\ww,
\frac{\Lambda^4}{\ww^2} \right),
\ee
and the  effective Lagrangian in the dual description will be given by
\be
{{\cal L}_{SW}}^D\!+\!{\cal L}_M\!+\!\int d^2\th d^2\bar{\th}
\;K(z,z^\dagger)
\!+\! \left( \int d^2 \th\ lz
\left\{\ww-U(A_D)+\frac{\Lambda^4}\ww f\left(\frac{A_D^2}\ww,
\frac{\Lambda^4}{\ww^2} \right)\right\}\! +\! h.c.\right).  
\label{wholag}
\ee
The elimination of the $F$-component of $z$ gives a potential term
\be
K_{z z^\dagger}^{-1}\;|l|^2
\left| \,\ww-u(a_D)+\frac{\Lambda^4}\ww 
f\left(\frac{a_D^2}\ww,\frac{\Lambda^4}{\ww^2} \right)\right|^2.
\label{potzet}
\ee
This term will of course complicate further analysis due to the 
unknown function $f$ and the unknown K\"{a}hler potential $K$.
We will {\em assume} that the auxiliary field limit of $z$, 
i.e.\ the limit $l\rightarrow\infty$, simplifies the analysis in such a
way 
that 
the potential \rf{potzet} simply provides a constraint which relates the 
free parameter $\ww$ and $a_D$.     
Since the parameter $\ww$ does not appear in the other terms of the
whole 
potential unless we consider higher derivative terms of the effective 
Lagrangian, we can regard $a_D$ as a free parameter instead of $\ww$,
and 
ignore the parameter $\ww$ in the following discussions.

\subsubsection{The vacuum structure}

We now analyze the vacuum structure of the softly broken model 
given by the Lagrangian \rf{wholag} with substitution of ${\cL_{SW}}^D$
with ${\cL_{soft}}^D$, following the analysis in \cite{ALV}.
Then, after the eliminations of the $D$ and $F$ components of the dual
photon
and the monopole fields, we obtain the potential
\bea
V_{soft+z}
&=&\frac1{b_{11}}\left|\, b_{01}\bar{F}_0+\sqrt{2}m\tilde{m}+zl
\frac{\pa u_q}{\pa a_D^1}\,
\right| ^2 
+\frac1{2b_{11}}\left( b_{01}D_0+|m|^2-|\tilde{m}|^2\right) ^2 \CR
&+&2|a_D|^2\,(|m|^2+|\tilde{m}|^2)-b_{00}|F_0|^2-\frac{b_{00}}2 D_0^2 
+\Bigl(F_0zl\frac{\pa u_q}{\pa a_D^0}+h.c.\Bigr),
\label{sofpot}  
\eea
where we ignored the potential (\ref{potzet}) for the reason mentioned
above.
In \rf{sofpot} $(m,\tilde{m})$ denote the scalar components
of the monopole hyper-multiplet $(M,\tilde{M})$ and 
\be
u_q=u-\frac{\Lambda^4}\ww f\left(\frac{a_D^2}\ww,\frac{\Lambda^4}{\ww^2} 
\right),
\ee
while 
\be
b_{ij}\equiv \frac{1}{4\pi}{\rm Im} \;\tau_{ij}
=\frac{1}{4\pi}{\rm Im}\;\frac{\pa^2{\cal F}}{\pa a_D^i\pa a_D^j}.
\ee
Finally $F_0$ and $D_0$ denote the frozen $F$ and $D$ components
of the spurion multiplet, respectively. 

In \cite{ALV} the choice of parameters 
\beq{*20}
F_0 \mbox{~\raisebox{-.8ex}{$\stackrel{\textstyle  <}{\sim}$}~}
\L,~~~~~D_0=0
\eeq 
was studied. In our case we find that this choice leads to 
an unbounded potential $V_{F_0}$ in a neighborhood of $a_D=0$. This 
implies that we have to take into account the contributions to the 
potential from higher derivative terms in order to understand 
the dynamics in this region. Since we have little control over 
these higher derivative terms we will not consider the case \rf{*20}
any further and turn to 
\beq{*21}
F_0 = 0,~~~~~D_0 \neq 0.
\eeq
{}From \rf{sofpot} we obtain, setting $F_0=0$,
\bea
V_{D_0}
&=&\frac1{b_{11}}\left|\sqrt{2}m\tilde{m}+zl\frac{\pa u_q}{\pa a_D^1}
\right| ^2 
+\frac1{2b_{11}}\left( b_{01}D_0+|m|^2-|\tilde{m}|^2\right) ^2 \CR
&+&2|a_D|^2(|m|^2+|\tilde{m}|^2)-\frac{b_{00}}2 D_0^2.
\label{potdze}
\eea
This potential has obviously a minimum with respect to the field $z$,
and we use this value for $z$.
Taking  derivatives with respect to the monopole fields, we obtain
\bea
\frac{\pa V_{D_0}}{\pa m}&=&\left( \frac1{b_{11}}(b_{01}D_0+|m|^2-
|\tilde{m}|^2)+2|a_D|^2\right)m=0, \CR
\frac{\pa V_{D_0}}{\pa \tilde{m}}&=&\left(
-\frac1{b_{11}}(b_{01}D_0+|m|^2-
|\tilde{m}|^2)+2|a_D|^2\right)\tilde{m}=0.
\label{monequ}
\eea
Thus if
\be
|a_D|^2<\frac{|b_{01}D_0|}{2b_{11}},
\label{regad}
\ee
these equations (\ref{monequ}) have solutions other than the trivial
$m=\tilde{m}=0$\footnote{If $D_0 =0$, we have $N \!\!=\!1$ and in
addition 
there is no region of monopole condensation. This is consistent with the 
holomorphy argument given below eq.\ \rf{potbarz}.}. They are
\bea
|\tilde{m}|^2&=&-2b_{11}|a_D|^2+b_{01}D_0,\ \ m=0 \ \ \ {\rm for~} 
b_{01}D_0>0, \CR 
|m|^2&=&-2b_{11}|a_D|^2-b_{01}D_0,\ \ \tilde{m}=0 \ \ \ {\rm for~} 
b_{01}D_0<0.
\label{solmmt}
\eea
In  both cases, $z=0$.
One can easily show that these non-trivial solutions correspond to 
the absolute minima. The non-zero vacuum expectation values of the 
monopole fields give a mass
to the dual photon. Following the general folklore this 
leads to a confined electric charge by  the dual Meissner effect.
Outside the region (\ref{regad}), the monopole fields do not condensate,
and the system has to be in the Coulomb phase.
Thus we have shown that by adding a term \rf{zetlag}
there may be a confinement-deconfinement phase transition 
line in the parameter space of $a_D$ (or $\ww$), {\em while the system
on 
both sides of the phase transition line can be considered as an
effective 
$N\!\!=\!0$ theory}. 

To regard this phase transition line as that of the pure compact $U(1)$,
it is important that no new massless degrees of freedom appear at the 
transition. As shown previously the system has 
a non-gauge global $U(1)\times U(1)_R$ symmetry\footnote{The breaking 
parameter $D_0$ has the $U(1)\times U(1)_R$ charge $D_0 =(0,0)$. }.
The $U(1)$ symmetry is anomalous, and hence is irrelevant here. 
The $U(1)_R$ symmetry might be broken in the Higgs phase of the dual 
description and thus generate zero-mass particles by the Goldstone
mechanism, 
but actually $U(1)_R$ is not broken because $m\tilde{m} 
= z =0$\footnote{Considering 
$m$ and $\tilde{m}$ separately, one may conclude there is a broken
symmetry.
But this is the dual $U(1)$ gauge symmetry and does not generate
massless
particles by the Higgs mechanism.}
for the solutions (\ref{solmmt}).
Hence, near the phase transition line, the light degrees of freedom are
the 
dual photon and the one of the monopole fields which becomes massless
at the phase transition line by  the Landau-Ginzburg mechanism.

In the next section we will analyze the phenomenology close to this 
phase transition line based on the picture of 
the dual photon and the monopole.
 
\section{Phenomenology}\label{sec4}

\subsection{Confinement phase}

In the confinement phase, close to  the phase boundary between 
the confinement and the deconfinement phase, we have an 
effective Ginzburg-Landau description in terms of monopole 
fields and a dual photon as in \rf{suppot} and \rf{suppot2}. 
Although the situation in \rf{potdze} and \rf{solmmt} 
is slightly more complicated than in the 
simplest breaking to $N\!\!=\!1$ described in 
\rf{suppot2} the order of magnitude 
is the same for the parameters involved,  and we will in the following 
use the simplified notation of a breaking parameter $b$ which 
monitors the distance from the points in coupling constant space 
where monopoles condense. In the original picture of 
Seiberg and Witten it corresponds to the point where $N\!\!=\!2$ is 
broken, which at the same time corresponds to the
confinement-deconfinement 
transition of the $U(1)$ charge. 
In the present generalized model $b$ corresponds to the distance to the 
confinement-deconfinement transition. 
In both cases we have (as in \rf{suppot2})
\beq{*31}
m \sim \sqrt {b \L_{N=2}}. 
\label{appm}
\eeq
In the more elaborate model considered above,  the 
breaking from $N=2$ to $N=0$ is partly separated from the 
confinement-deconfinement transition. The supersymmetry breaking 
occurs at a scale $D_0$, while monopole condensation
is dictated by \rf{regad}. Thus 
$b$ is here a function of $D_0$ and $a_D$ (or $\ww$), as will be
discussed 
below.

Let us now relate the scale $b$ to the non-perturbative 
physical scale in the confinement phase. Consider first the 
simplest case where $N\!\!=\!2$ is broken to $N\!\!=\!1$.
The running coupling constant is given by 
\beq{*32}
\frac{1}{g^{2}_{N=2}(\m)} = \frac{4}{8\pi^2} \log \frac{\m}{\L_{N=2}}.
\eeq 
Now assume that the breaking scale $b$ is larger than the dynamical
scale
$\L_{N=2}$. When the theory is broken to $N\!\!=\!1$, some of the
components
become massive and will not contribute to the further running 
of the coupling constant. For the light $N\!\!=\!1$ components we have
\beq{*33}
\frac{1}{g^{2}_{N=1}(\m)} = \frac{6}{8\pi^2} \log \frac{\m}{\L_{N=1}}.
\eeq
The scales of the theories can be related by the matching condition
$g^2_{N=1}(b) = g^2_{N=2} (b)$ at the breaking scale, which, according 
to standard arguments, implies \cite{FINPOU} 
\beq{*34}
\L_{N=1}^6 = b^2 \L^4_{N=2}.
\eeq

We can repeat the argument in the case where the breaking is all the 
way from $N\!\!=\!2$ down to $N\!\!=\!0$. 
Now the parameter $b$ includes two free 
parameters $D_0$ and $a_D$ (or $w$) as can be seen from \rf{solmmt} and 
\rf{*31}:
$
b = (|b_{01} D_0| -2b_{11} |a_D|^2)/{\L_{N=2}}.
$
Since $D_0$ is the breaking scale of $N\!\!=\!2$ supersymmetry, 
it has a clearer 
meaning in the original system than the parameter $a_D$ (or $w$) which  
is linked to the unknown function $f$ and 
the limit $l\rightarrow \infty$ in \rf{potzet}.    
Thus, in the following discussion, 
for the purpose of clarity let us 
leave the parameter $a_D$(or $w$) fixed and regard the 
$D_0$ as the free parameter which monitors the phase transition.
Assume now that the parameter $D_0$ is larger than the dynamical scale 
$\L_{N=2}$. Then comparing \rf{solmmt} and \rf{appm}, the parameter 
$b$ may be identified to the breaking scale $D_0$, since 
$b_{01}\sim \L_{N=2}$ and $a_D$ is considered fixed.
In this  case the fermionic partner to the gauge
field becomes massive below the breaking scale and \rf{*33} is 
replaced by 
\beq{*35}
\frac{1}{g^2_{N=0} (\m)} = \frac{11}{12\pi^2} \log  \frac{\m}{\L_{N=0}},
\eeq
{}from which we conclude that 
\beq{*36}
\L_{N=0}^{\frac{22}{3}} = b^{\frac{10}{3}} \L^4_{N=2}.
\eeq
The non-perturbative scales $\L_{N=1}$ and $\L_{N=0}$, respectively, 
are the scales in which any mass  (string tension, gauge-balls, etc.)
should 
be measured in the confinement regime of the $N\!\!=\!1$ 
and the $N \!\!=\!0$ theories.

\subsubsection{Mapping on compact lattice U(1)}

If we want to map the physics of the $U(1)$ part of the 
broken $N\!\!=\!2$ supersym\-me\-tric theory onto the 
compact $U(1)$ lattice theory 
it is natural to identify the symmetry breaking parameter $b$ 
with $\b_c -\b$, since one in this case gets a Ginzburg-Landau
expression
for the monopole condensate from \rf{*31} and the  
monopole condensate forms on the lattice when $\b_c-\b$ becomes positive 
and it forms in the continuum model when $b >0$. 
Under this {\em assumption}, plus 
the {\em assumption} that the qualitative physics of the $U(1)$ 
sector is the same in the 
continuum theory and the lattice theory (like in three dimensions), 
we get for the non-perturbative mass scales $\L(\b)$ in the confinement 
sector:
\beq{*37}
\L(\b) \sim  (\b_c -\b)^{\frac13}~~~{\rm for}~~~N\!\!=\!2 \to N\!\!=\!1,
\eeq
\beq{*38}
\L(\b) \sim  (\b_c -\b)^{\frac5{11}}~~~{\rm for}~~~N\!\!=\!2 \to
N\!\!=\!0.
\eeq
These exponents are just the ones measured in the lattice simulations of 
various compact $U(1)$ theories, as discussed in sec.\ \ref{sec2}.
Further, the above scenario also explains why there should be 
excitations in the system associated with the mean field 
Ginzburg-Landau exponents
\beq{*39}
M(\b) \sim (\b_c-\b)^{\frac12}
\eeq
for excitations of the monopole condensate.

We do not pretend that the arguments above are conclusive. 
Strictly speaking the matching of scales by the one loop formula is 
only valid if $b$ is large, and we want to use the matching in a region 
where it is not the case. Further, the breaking to $N\!\!=\!1$ is of
course 
an incomplete breaking, and if the mass exponent $1/3$ should be 
compared to lattice results, further breaking down to $N\!\!=\!0$ should 
be implemented in a way not affecting the scaling. We have no 
explicit suggestions to how this could occur. In addition the 
actual value of the renormalized charge fine-structure constant 
$\a$ in the lattice simulations is $\a \approx 0.2$, i.e.\ $\a_D \approx
5$.
This is not exactly in the region where the dual photon--monopole
picture 
is believed to be reliable. On the other hand the direct 
translation of the lattice charge $\a$ to the continuum is not clear, so
these
values might be misleading from a continuum point of view.
Finally, the occurrence
of a non-perturbative scaling and a mean field scaling at the 
same time is a somewhat unusual situation. Nevertheless, it is 
precisely what {\em is} observed in the lattice simulations.
{}From the above considerations 
the physics of the two different lattice systems 
studied ($U(1)$ without an explicit monopole term in the action, and 
$U(1)$ with a monopole term included) should then be identified with the 
physics of different softly broken $N\!\!=\!2$ supersymmetric theories
and this identification explains to some extent the observed non-trivial
critical exponents ($\n = 0.35\pm 0.03$ and $\n=0.44\pm 0.02$) 
and in addition the mean field exponents which are also observed.

\subsubsection{The one-particle spectrum}

In order to substantiate further that the physics
of compact lattice $U(1)$ is the same as broken $N\!\!=\!2$
supersymmetric 
Yang-Mills theory we discuss below  
the parity and charge conjugation quantum numbers of 
the one-particle state in the confinement phase of the broken 
supersymmetric theory and compare it with the measurements on the
lattice.

The pure $U(1)$ gauge theory $\int d^4x F_{\mu\nu}F^{\mu\nu}$
is obviously invariant under the parity and charge conjugation
transformation
defined by 
\bea
P&:&\ A_0\rightarrow A_0,\ \ A_i\rightarrow -A_i, \CR
C&:&\ A_\mu\rightarrow -A_\mu,
\eea
where $A_\mu$ are the gauge potential.
Under these transformations the field strengths transform as
\bea
P&:&\ E\rightarrow -E,\ \ B\rightarrow B, \CR
C&:&\ E,B\rightarrow -E,-B. 
\eea
Defining the duality transformation by $E^D=B$ and $B^D=E$,
one obtains the parity and charge conjugation transformation of the 
dual gauge field as
\bea
P&:&\   A_0^D\rightarrow -A_0^D ,\ \ A_i^D\rightarrow A_i^D , \CR
C&:&\ A_\mu^D\rightarrow -A_\mu^D,
\eea
Thus the dual photon field is $1^{+-}$, where $J^{PC}$ denotes a spin
$J$,
parity $P$ and charge conjugation $C$ state.
Since the covariant derivative $D_\mu^D\equiv \pa_\mu+iA^D_\mu$
transforms as
\bea
P&:&\ D_0^D\rightarrow {D_0^D}^\star,\ \ D_i^D\rightarrow
-{D_i^D}^\star, \CR
C&:&\ D_\mu^D\rightarrow {D_\mu^D}^\star,
\eea
the transformations of the monopole field $M$ are given by
\bea
P&:&\ M\rightarrow \eta_P~ M^\star, \CR
C&:&\ M\rightarrow \eta_C~ M^\star, 
\eea
where $\eta_{P,C}$ are the phase factor ambiguities coming from the
global 
$U(1)$ dual gauge symmetry of the monopole field action.
To fix this ambiguity, we take the vacuum expectation value $v_M$ of the 
monopole field to be real by using the global gauge transformation. 
Then the appropriate $P$ and $C$ transformations
which preserve the vacuum expectation value are given by just taking 
$\eta_{P,C}=1.$\footnote{In fact one can continue the discussion without 
taking special values of $\eta_{P,C}$. In this case one must consider 
the gauge invariant field $M_R\equiv {v_M}^\star M$ instead of just 
$M_R\equiv {\rm real\ part\ of}\ M$ as the physical monopole field. 
For simplicity, we take the special values.}
Thus one concludes that the physical monopole field, 
$M_R\equiv {\rm real\ part\ of\ }M$, has the quantum number $0^{++}$.

This quantum number assignment is quite interesting. 
The lattice spectroscopy shows that there are the one-particle
states with quantum numbers $0^{++}$ and $1^{+-}$ in the confinement
phase \cite{NEU}. We identify these states as the monopole
(or, in the case where a monopole condensate is formed, as the lowest
excitation in this condensate) and the dual
photon one-particle state, respectively. We see that this assignment is 
in accordance with the identification of the monopole condensate
excitations as
the ones with the mean field exponents, since the $0^{++}$ state 
has $\n =1/2$.

\subsection{Coulomb phase}

In the broken supersymmetric model we move into 
the Coulomb phase when $a_D$ is sufficiently large. Deep into this phase 
the low energy fields excitations involve just the ordinary photon
field.
The monopoles will be very heavy. This is trivially in agreement with 
the lattice picture. As mentioned in sec.\ \ref{sec2} the physics 
of compact lattice $U(1)$ is quite interesting near the phase
transition:
There seems to be a universal value of the ``renormalized'' charge 
in the theory, caused by the anti-screening of the vacuum 
monopole fluctuations. From the philosophy used so far, one could 
hope that it was possible to predict the value $\a_R^c$, as well as the 
critical exponent $\l$ in the expression \rf{*2}.
We have not yet made any progress in that direction.   

\section{Summary and discussions}\label{sec5}

The recent remarkable lattice simulations of compact $U(1)$ show
very  interesting physics with a confinement-deconfinement 
transition which is reported to be of  second order 
with non-trivial critical mass exponents.

{\it If} the transition is  second order, we have a new and very 
interesting situation in quantum field theory since 
there should be an underlying continuum field theory of a new kind. 
Presently we have nothing more to say about such a revolutionary 
scenario. 

In this paper we have tried something more modest, namely to 
find {\em a} continuum, four-dimensional field theory which at a 
qualitative level has the same low energy physics as the lattice model
and thus might explain the observed critical exponents.
Thus this continuum model will retain its explanatory value 
even if the transition is not really a second order 
transition, but a weakly first order transition which has  only 
pseudo-critical exponents. 

The $XY$-model and the sine-Gordon model (at a particular value 
of the coupling constant) constitute an  example 
of the first kind: a non-trivial
scaling limit of the $XY$-model exists and the physics in this 
limit is identical to that of the sine-Gordon theory at a special 
value of the coupling constant. Three-dimensional compact lattice $U(1)$
and the Georgi-Glashow model provides an  example 
of the second kind: the physics
is qualitatively of the same nature for low energy excitations, but 
no genuine scaling limit can be defined for the lattice theory, such
that 
it becomes identical to the Georgi-Glashow model. 
According to folklore (which might now be falsified by the 
$U(1)$ lattice models) the supersymmetric
field theories are the only four-dimensional 
field theories containing  scalar fields, which have chance to  
exist as non-trivial, genuine interacting field theories.
We have analyzed the soft breaking of a class of such continuum 
theories and have shown that the non-trivial dynamics 
of these theories indeed might explain the observed critical or 
pseudo-critical exponents, as well as the discrete quantum numbers 
assigned to some of the particles in the lattice theory. 

If one assumes our heuristic arguments are correct, there is  
a chance that one in the future will be able to understand 
why the renormalized lattice charge in the Coulomb phase seems to 
converge 
to a universal number independent of the details of lattice
implementation of the theory for $\b \to \b_c$.

\section*{Acknowledgements}

D.E. and N.S. would like to thank M.~Mari\~no for explaining the  
results reported in \cite{ALV}. D.E. would like to thank L.
Alvarez-Gaum\'e and 
F. Zamora for early discussions on the subject and the CERN TH-Division 
for hospitality. In addition we thank
P. Di Vecchia, J. L. Petersen, J. Jersak,
C. Lang, T. Neuhaus and P. Igor
for reading the article prior to publication and for many useful
comments.
N.S. is supported by DANVIS grant No.~1996-145-0003
{}from the Danish Research Academy. D.E. acknowledges the support
{}from CICYT grant AEN95-05090-0695 and CIRIT contract
GRQ93-1047.
J.A. acknowledges the support of the Professor Visitante Iberdrola 
Program and the hospitality at the University of Barcelona, where part
of this work was done.

\end{document}